\documentclass[reprint,
 superscriptaddress,
 amsmath,amssymb,
 aps,
 prx,
]{revtex4-2}

\usepackage{graphicx}
\usepackage{amsmath}
\usepackage{mathtools}
\usepackage{comment}
\usepackage{bbm}
\usepackage{lipsum}
\usepackage{amssymb}
\usepackage{xcolor}
\usepackage{caption}
\usepackage{capt-of}
\definecolor{red}{rgb}{1,0.,0}

\usepackage{graphicx}
\usepackage{subcaption}
\usepackage{dcolumn}
\usepackage{bm}

\begin{document}


\title{Understanding multiscale disorder in superconducting nanowire single photon detectors}

\author{Nirjhar Sarkar}
\email{nirjharize@gmail.com}
\affiliation{Materials Science and Technology Division, Oak Ridge National Laboratory, Oak Ridge TN 37830, USA}
\author{Ronan Gourgues}
\affiliation{Single Quantum B.V., Delft, the Netherlands}
\author{Yueh-Chun Wu}
\affiliation{Materials Science and Technology Division, Oak Ridge National Laboratory, Oak Ridge TN 37830, USA}
\author{Chengyun Hua}
\affiliation{Materials Science and Technology Division, Oak Ridge National Laboratory, Oak Ridge TN 37830, USA}
\author{Katyayani Seal}
\affiliation{Single Quantum B.V., Delft, the Netherlands}
\author{Andreas Fognini}
\affiliation{Single Quantum B.V., Delft, the Netherlands}
\author{Steven Randolph}
\affiliation{Center for Nanophase Materials Sciences, Oak Ridge National Laboratory, Oak Ridge TN 37830, USA}
\author{Eugene Dumitrescu}
\affiliation{Computational Science and Engineering Division, Oak Ridge National Laboratory, Oak Ridge TN 37830, USA}
\author{G\'abor B. Hal\'asz}
\affiliation{Materials Science and Technology Division, Oak Ridge National Laboratory, Oak Ridge TN 37830, USA}
\author{Benjamin Lawrie}
\email{lawriebj@ornl.gov; This manuscript has been authored by UT-Battelle, LLC, under contract DE-AC05-00OR22725 with the US Department of Energy (DOE). The US government retains and the publisher, by accepting the article for publication, acknowledges that the US government retains a nonexclusive, paid-up, irrevocable, worldwide license to publish or reproduce the published form of this manuscript, or allow others to do so, for US government purposes. DOE will provide public access to these results of federally sponsored research in accordance with the DOE Public Access Plan (http://energy.gov/downloads/doe-public-access-plan). }
\affiliation{Materials Science and Technology Division, Oak Ridge National Laboratory, Oak Ridge TN 37830, USA}
\date{\today}

\begin{abstract}
Superconducting nanowire single-photon detectors are central to applications across quantum information science. Yet, their performance is limited by the effects of disorder and electrodynamic inhomogeneities that are not well understood. By combining DC transport, dark-count measurements, and bias-dependent microwave transmission spectroscopy in the presence of controlled nanoscale disorder introduced through helium-ion irradiation, we distinguish local instability-driven processes from intrinsic superconducting depairing and kinetic inductance nonlinearities. This approach enables systematic tuning of kinetic inductance, depairing currents, microwave dissipation, and mode structure within a single device. Bias- and temperature-dependent resonance shifts quantify disorder-induced modifications of the superconducting density of states through the nonlinear kinetic inductance, while the emergence of multiple resonant modes reveals the formation of electrodynamically distinct superconducting regions. Comparing depairing under current, field, and temperature isolates the dominant microwave loss mechanisms, separating vortex, quasiparticle, and two-level-system contributions, thus providing a robust multifunctional foundation for disorder engineering of superconducting nanowire detectors and resonators.

\end{abstract}

\pacs{Valid PACS appear here}

\maketitle

Superconducting nanowire single-photon detector (SNSPD) performance is strongly dependent on disorder introduced during thin-film growth and nanofabrication, which produces highly localized weak spots that behave like constrictions\cite{kerman2007constriction,clem2011geometry,semenov2015asymmetry}. These constrictions can clamp the device’s usable bias current density well below the intrinsic uniform limit (the depairing current $I_{\mathrm{dep}}$). Helium ion irradiation has emerged as a controlled, post-fabrication approach to locally tune this disorder landscape in SNSPDs without modifying device geometry. Indeed, prior work has demonstrated control over switching current, detection efficiency, hotspot relaxation dynamics, and substrate-mediated thermal transport following irradiation \cite{strohauer2025current,hong2025impact,zhang2019saturating}.

However, the impact of helium-induced disorder on the microwave electrodynamics of SNSPDs has not been systematically explored. Microwave spectroscopy provides direct access to the kinetic inductance, dissipation, and depairing in superconducting nanowires\cite{frasca_determining_2019,santavicca2016microwave} and thus offers a powerful tool for probing the effects of locally patterned disorder in superconducting nanowires. Disorder plays different roles across applications. In circuit-QED and high-impedance resonators, it is often introduced deliberately to increase kinetic inductance and impedance at the cost of increased microwave loss and reduced dynamic range \cite{maleeva2018circuit}. In photon detection, the same disorder improves detection efficiency but lowers the usable current density and degrades reset time and timing jitter\cite{haldar2025influence,strohauer2025current}.

\begin{figure}[hbt!]
\centering
    \includegraphics[width=0.6\columnwidth]{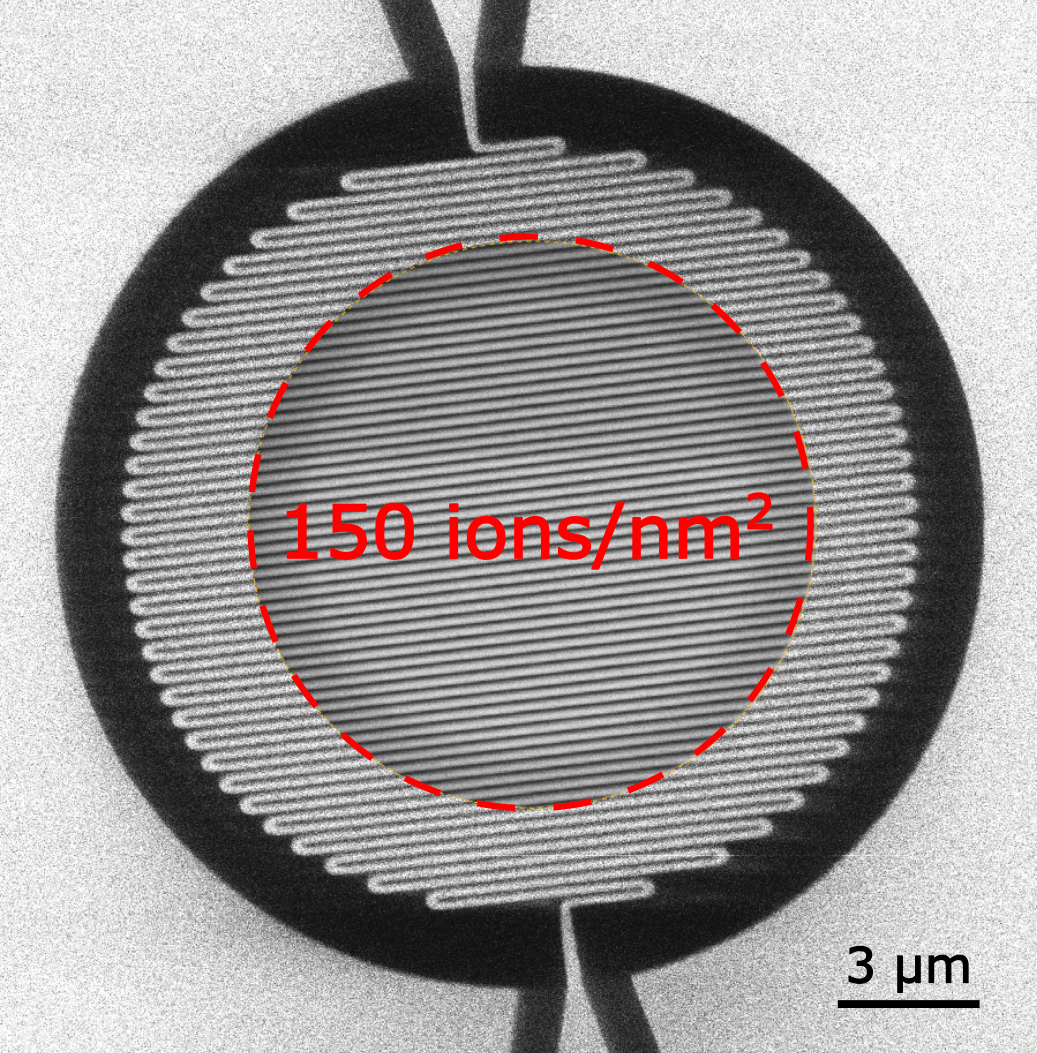}
    \caption{Helium ion microscope image of a NbTiN SNSPD with the locally irradiated region indicated by a red dashed circle. The ion fluences used in this work are 50 and 150 ions/nm$^2$. All ion implantation was performed at 30 keV.}
    \label{fig:fig1}
\end{figure}

\begin{figure*}[hbt!]
\centering
    \includegraphics[width=\textwidth]{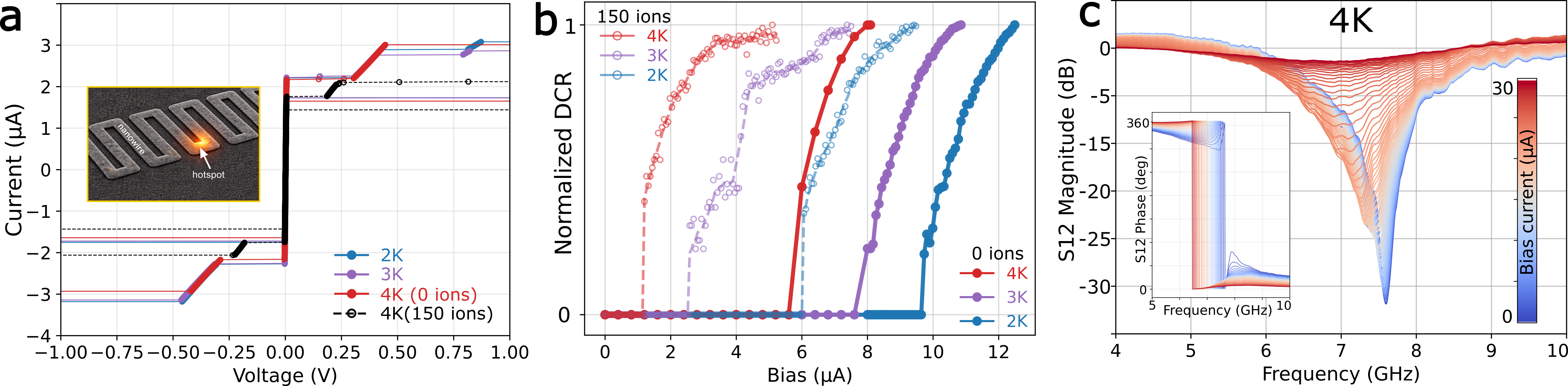}
    \caption{Three characteristic current scales for the same SNSPD extracted from different measurements. (a) DC I–V characteristics at 2 K (blue), 3 K (purple), and 4 K (red), showing a first local switching event near $\approx2~\mu$A; the dashed black curve is the 2 K I–V measurement after helium-ion implantation ($150~\mathrm{ions}/\mathrm{nm}^{2}$). The left inset illustrates a hotspot at nanowire bends near $I_{\mathrm{SW}}$. (b) Dark-count rate versus bias current for the same temperatures and color scheme: solid curves (non-irradiated) show onsets at $\approx$ 10, 8, and 6 $\mu$A, while dashed, semi-transparent curves (locally disordered $150~\mathrm{ions}/\mathrm{nm}^{2}$) show reduced onsets at $\approx$ 6, 3, and 1 $\mu$A. (c) Microwave transmission $|S_{12}|$ versus frequency at 4 K for different bias currents. The resonant peak shifts to a lower frequency and becomes strongly damped near $\approx25~\mu$A. The highly damped resonance was tracked clearly in the phase response shown in the left inset. Together, these 3 measurements probe distinct aspects of the device physics. }
    \label{fig:fig2}
\end{figure*}

In this work, we show that the intrinsic depairing current $I_{\mathrm{dep}}$ in NbTiN SNSPDs is more strongly affected by helium-ion-induced disorder than the limiting currents extracted from standard SNSPD performance metrics. The devices used here are patterned from a 10 nm NbTiN thin film on a distributed Bragg reflector optimized for telecom wavelengths. The patterned nanowire meander pattern has a wire width of 100 nm and a pitch of 200 nm. Local-area helium ion irradiation is applied to up to $80\%$ of the device area with He$^{+}$ fluences of up to $150~\mathrm{ions}/\mathrm{nm}^{2}$ on the straight segments of the meandering nanowires (to avoid current crowding at bends \cite{strohauer2025current}), as shown in Fig.~\ref{fig:fig1}. 

In practice, SNSPD current limits are obtained from current-voltage (IV) measurements and dark-count measurements, which probe different dissipation regimes of the same DC-biased nanowire. The dark count rate (DCR) increases exponentially at a threshold current ($I_{\mathrm{DCR}}$) and corresponds to self-resetting dissipative events, modeled as vortex crossings or phase-slip-like processes. Every reset generates transient voltage pulses that register as a dark count while the wire remains superconducting between events. In contrast, for an IV measurement, the first local switching current ($I_{\mathrm{SW}}$) is the bias at which a self-heating normal domain nucleates at the weakest section of the wire (typically constrictions and bends) and becomes electrothermally stable, latching the device into a persistent resistive state even though the remainder of the nanowire may remain superconducting. Dark counts are not observed at $I_{\mathrm{SW}}$ because thermal latching eliminates self-resetting dynamics. 

To understand the effect of helium ion implantation, a combination of IV, DCR, and RF spectroscopy are used as a function of varying irradiation dose, temperature, magnetic field, and bias current to probe different ``limiting physics,'' as shown in Figure~\ref{fig:fig2}. This combination of measurements provides a complete picture of the effects of controllably patterned disorder across length scales.

Figure~2a shows the first local switching current ($I_{\mathrm{SW}}$) near $2~\mu\mathrm{A}$ at $4~\mathrm{K}$, extracted from IV sweeps that are slowly time-averaged over seconds. Figure~2b shows the dark-count onset current ($I_{\mathrm{DCR}}$) near $10~\mu\mathrm{A}$ at $4~\mathrm{K}$, where the RF port is monitored through a bias-tee and RF amplifier to count nanosecond-scale voltage pulses. Figure~2c shows the upper intrinsic limit, the depairing current ($I_{\mathrm{DEP}}$), near $25~\mu\mathrm{A}$ at $4~\mathrm{K}$. The depairing current is obtained by tracking the superfluid response via the microwave transmission spectrum $S_{21}$ while biasing the device through the same bias tee; the resonance peak at frequency $f_0$ shifts and becomes strongly damped near $25~\mu\mathrm{A}$. In this regime, the resonance frequency is more reliably identified from the phase response, which continues to show a sharp transition centered at $f_0$ even when the $S_{21}$ magnitude reduces. These measurements establish the ordering
\[
I_{\mathrm{SW}} < I_{\mathrm{DCR}} < I_{\mathrm{DEP}} .
\]

The dashed curves in Figs. 2a and 2b show changes in the IV sweeps and DCR measurements after helium-ion implantation at dose of 150 ions/nm$^2$. $I_{\mathrm{SW}}$ and $I_{\mathrm{DCR}}$ change from $(2,\,10)~\mu\mathrm{A}$ to $(2,\,6)~\mu\mathrm{A}$ at $4~\mathrm{K}$. The negligible change in $I_{\mathrm{SW}}$ at this implantation dose indicates that the DC switching event likely originates outside the irradiated region, most plausibly at current-crowded bends; higher irradiation doses would be expected to suppress $I_{\mathrm{SW}}$.

The reduced $I_{\mathrm{DCR}}$ reflects disorder-induced modification of the straight nanowire segment. After ion irradiation, the normalized DCR curves in Fig. 2b showed an earlier onset with a broader DCR turn-on, indicating that disorder introduces a distribution of multiple vortex-crossing sites rather than a single bottleneck activating over different bias currents. The irradiated straight segment therefore dominates the dark-count response and limits the usable bias window after irradiation. Together, $I_{\mathrm{SW}}$ and $I_{\mathrm{DCR}}$ probe local instability thresholds associated with geometric or microscopic weak spots, whereas microwave spectroscopy accesses the global superconducting response averaged over the nanowire. Among these, $I_{\mathrm{DEP}}$ shows the strongest and most systematic suppression with increasing disorder, motivating a detailed microwave-spectroscopy investigation via controlled ion implantation in this work.

\begin{figure}[t]
\centering
    \includegraphics[width=\columnwidth]{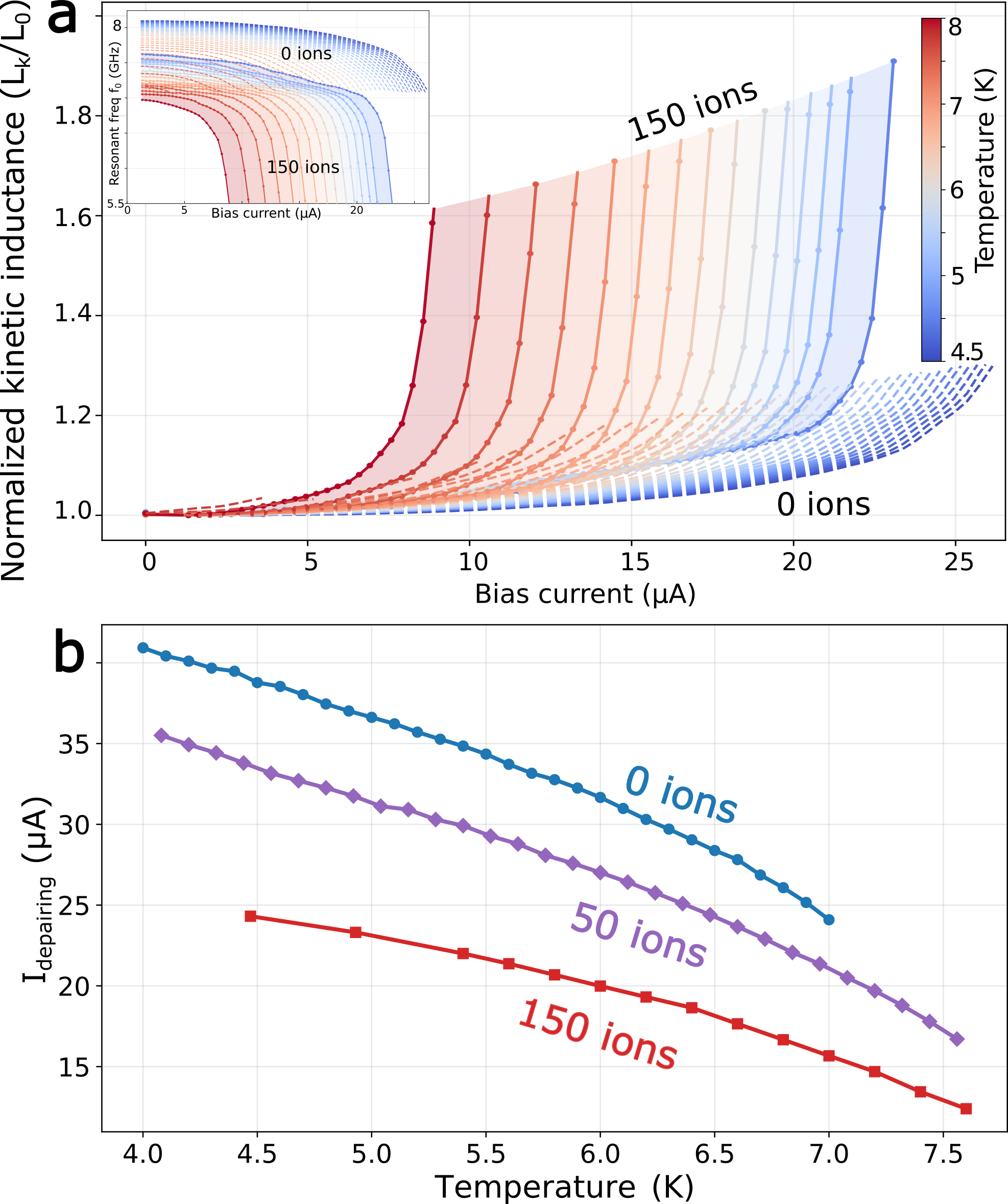}
    \caption{Bias- and temperature-dependent kinetic inductance of disordered SNSPDs. (a) Normalized kinetic inductance $L_k(I,T)/L_k(0,T)$ extracted from the resonance frequency $f_0(I,T)$ shown in the inset, using $L_k \propto 1/f_0^2$. (b) Extracted depairing current versus temperature obtained by fitting the data in (a) using the Clem--Kogan fast-relaxation model for SNSPDs with different irradiation fluences.}
    \label{fig:fig3}
\end{figure}

In contrast to the modest changes observed in $I_{\mathrm{SW}}$ and $I_{\mathrm{DCR}}$, the depairing current extracted from microwave spectroscopy shows the strongest and most systematic suppression with increasing disorder, accompanied by an enhanced kinetic inductance quantified by the bias-induced resonance-frequency shift as shown in Figure 3. 

Figure~\ref{fig:fig3} quantifies how increasing ion implantation doses enhance the global kinetic inductance of the nanowire and suppress the intrinsic depairing current. The inset of Fig.~\ref{fig:fig3}a illustrates the resonance frequency \(f_0\) as a function of bias current for non-irradiated and irradiated devices (at 150 ions/nm$^2$), showing a larger downshift of frequency with bias for the irradiated device. Using $L_k \propto 1/f_0^2$, Fig.~\ref{fig:fig3}(a) shows the extracted normalized kinetic inductance \(L_k(I)/L_{k,0}\) as a function of bias current for both devices. The non-irradiated nanowire shows only a weak bias dependence of $L_k$, whereas irradiated devices exhibit an earlier and steeper increase, indicating that the disordered region reaches its intrinsic instability at lower currents. Fitting the full $L_k(I,T)$ curves with the Clem--Kogan fast-relaxation model yields $I_{\mathrm{dep}}(T)$ for each ion fluence, as shown in Fig.~3b for the pristine device and devices irradiated at $50~\mathrm{ions}/\mathrm{nm}^{2}$ and $150~\mathrm{ions}/\mathrm{nm}^{2}$, providing a global depairing scale that is insensitive to isolated geometric bottlenecks and instead reflects the intrinsic electrodynamics of the composite nanowire~\cite{clem2012kinetic}.

\begin{figure*}[t]
\centering
    \includegraphics[width=\textwidth]{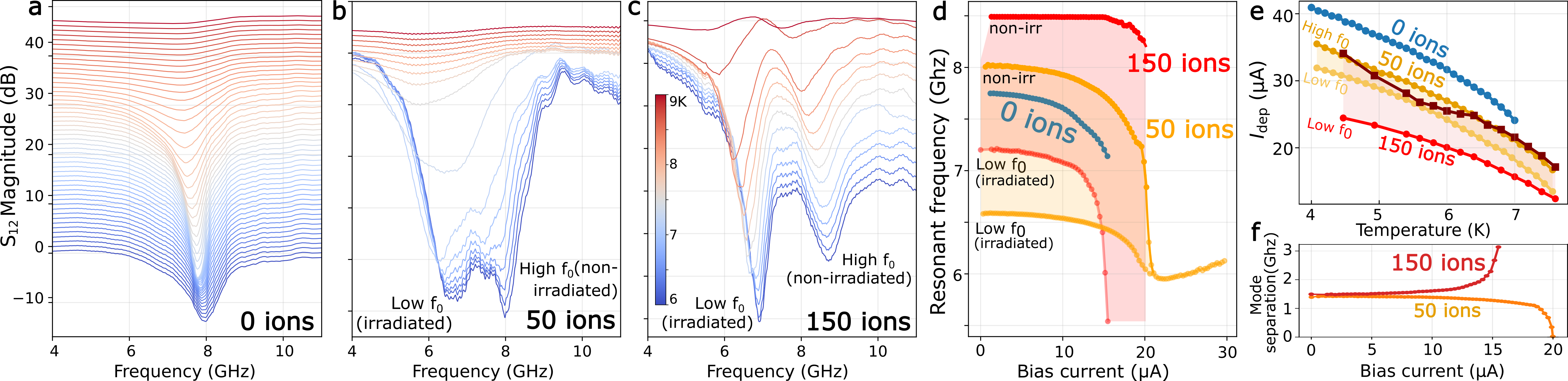}
    \caption{Local helium ion irradiation creates two superconducting regions with distinct microwave resonances. Microwave transmission spectra $|S_{12}|$ for devices irradiated at 0, 50, and 150 ions/nm$^{2}$ in (a), (b), and (c) for variable temperature illustrated by the shared color bar in (c). Increasing helium ion doses progressively broaden and then split the resonance into two distinct modes. (d) Bias-dependent resonance frequencies extracted from the microwave spectra for all irradiation doses. The 150 ions/nm$^{2}$ device exhibits two well-separated branches across the full bias range, while the 50 ions/nm$^{2}$ device shows bias-induced convergence and merging of the two branches. (e) Extracted depairing current versus temperature for each branch, indicating that the low-frequency resonance has a lower depairing current corresponding to the disordered superconducting region. (f) Resonance detuning $\Delta f_{\mathrm{meas}}(I)=f_{\mathrm{hi}}-f_{\mathrm{lo}}$ extracted for both 150 and 50 ions. }
    \label{fig:fig4}
\end{figure*}
The bias dependence of the kinetic inductance provides a quantitative probe of microscopic pair breaking in disordered superconducting nanowires. In the dirty limit relevant here~\cite{kubo2025higgs}, the kinetic inductance can be approximated in the low bias-current limit as
\[
\frac{L_k(I_b)}{L_k(0)} = 1 + C\left(\frac{I_b}{I_{\mathrm{dep}}}\right)^2 + \ldots,
\]
where the coefficient $C$ parameterizes the strength of the nonlinear Meissner response and is determined by the quasiparticle density of states. For a clean BCS thin film, $C\approx0.4$ as predicted by Kubo's dirty-limit theory~\cite{kubo2025higgs}, whereas strong disorder or subgap states suppress $C$ toward $\sim0.1$. From fits to $L_k(I)$, we extract $C\approx0.30$ for the non-irradiated device and $C\approx0.12$ for the 150~ions/nm$^{2}$ device (see Supplementary Information). The reduced value in the non-irradiated device relative to the clean-limit expectation indicates moderate intrinsic disorder arising from material growth and nanowire fabrication. Helium-ion irradiation further enhances this disorder, driving \(C\) toward the dirty-limit regime. From Kubo--Dynes theory, the measured reduction of $C$ corresponds to an increased effective $\Gamma/\Delta$, which describes disorder-induced broadening of the superconducting density of states, providing a device-level measure of disorder~\cite{kubo2025higgs}. An independent determination of \(\Gamma/\Delta\) by local tunneling spectroscopy could offer a direct and local microscopic validation of the disorder parameter $C$ inferred from microwave device measurements in future work~\cite{chen2023visualizing}. 
Helium-ion disorder provides a simple post-fabrication knob to tune the nonlinear kinetic-inductance coefficient C, enabling the optimization of nonlinear resonators suitable for parametric amplification and tunable circuit-QED elements~\cite{maleeva2018circuit,samkharadze2016high}, as well as weakly nonlinear regimes favorable for microwave kinetic inductance detectors (MKIDs) and superinductors~\cite{gao2008experimental}.

Figure~\ref{fig:fig4}(a-c) illustrates the measured microwave transmission spectra $|S_{12}|$ for a pristine SNSPD along with SNSPDs irradiated at 50 and 150 ions/nm$^{2}$. While the non-irradiated device exhibits a single resonance, the irradiated devices develop two distinct resonances that persist and evolve systematically with temperature. Microscopically, the meandering current path traverses many short alternating irradiated and non-irradiated segments but electrodynamically only two resonances are observed. The modulation length scale of the irradiated region ($\approx 12~\mu\mathrm{m}$) is orders of magnitude smaller than the microwave's electrical wavelength at 6–9 GHz ($\approx$$2-5~\mathrm{cm}$). As a result, the resonator responds to the irradiated area as a single distributed kinetic-inductor rather than discrete elements, consistent with distributed-parameter transmission lines \cite{pozar2021microwave}.

Bias- and temperature-dependent tracking of each resonance allows independent extraction of $f_0$, $I_{\mathrm{dep}}$ and $\Delta f_{\mathrm{meas}}$ in Fig.~\ref{fig:fig4}(d--f). In Fig. 4e, the $I_{\mathrm{dep}}$ extracted from the high-frequency resonances of the 50 and 150 ions/nm$^{2}$ devices overlap, suggesting that these modes are associated with the non-irradiated segments. In contrast, the low-frequency branches are separated, with the 150-ions/nm$^{2}$ device exhibiting a lower $I_{\mathrm{dep}}$ than the 50-ions/nm$^{2}$ device, suggesting that this branch originates from the irradiated segments. This is also consistent with $L_k \propto 1/f_0^{2}$ whereby the lower- (higher-) frequency resonance corresponds to larger (smaller) kinetic inductance in the irradiated (non-irradiated) region.

Figures 4d and 4f show the bias-current evolution of the two resonance frequencies and their measured splitting, defined as $\Delta f_{\mathrm{meas}}(I)=f_{\mathrm{hi}}-f_{\mathrm{lo}}$. In the strongly disordered device (150 ions/nm$^{2}$), the two branches remain well separated over the full bias range. $\Delta f_{\mathrm{meas}}$ increases with bias, indicating large and growing detuning between irradiated and non-irradiated sections as the irradiated segment enters a nonlinear kinetic-inductance regime at lower current (see Fig. 2a). In contrast, the intermediate-disorder device (50 ions/nm$^{2}$) shows bias-induced convergence of the two branches, yielding an effectively single-impedance response at high bias current. This contrasting behavior identifies an intermediate-disorder crossover in which DC bias tunes the system from a coupled two-mode response at low bias to a single-mode electrodynamic response at high bias, with $\Delta f_{\mathrm{meas}}$ providing a direct metric of disorder-induced inductance contrast.

Notably, at the highest fluence ($150~\mathrm{ions}/\mathrm{nm}^{2}$), the low-frequency resonance associated with the irradiated region becomes sharper than the high-frequency resonance associated with the non-irradiated region. This indicates that the two modes are limited by different dissipation mechanisms. The high-frequency mode corresponding to the non-irradiated section shows increased microwave loss, possibly because it contains bends and current-crowding regions where locally enhanced current density increases dissipative processes and broadens the resonance.

Beyond SNSPDs,  precise control of resonance frequencies and spacing is required in large MKID arrays to avoid frequency collisions and maximize multiplexing yield\cite{liu2017superconducting}. Controlled helium-ion implantation can be used to locally modify the kinetic inductance, enabling deterministic frequency shifts and mode-dependent quality-factor tuning as a post-fabrication strategy to compensate for kinetic-inductance nonuniformity. 

\begin{figure*}[t]
\centering
    \includegraphics[width=\textwidth]{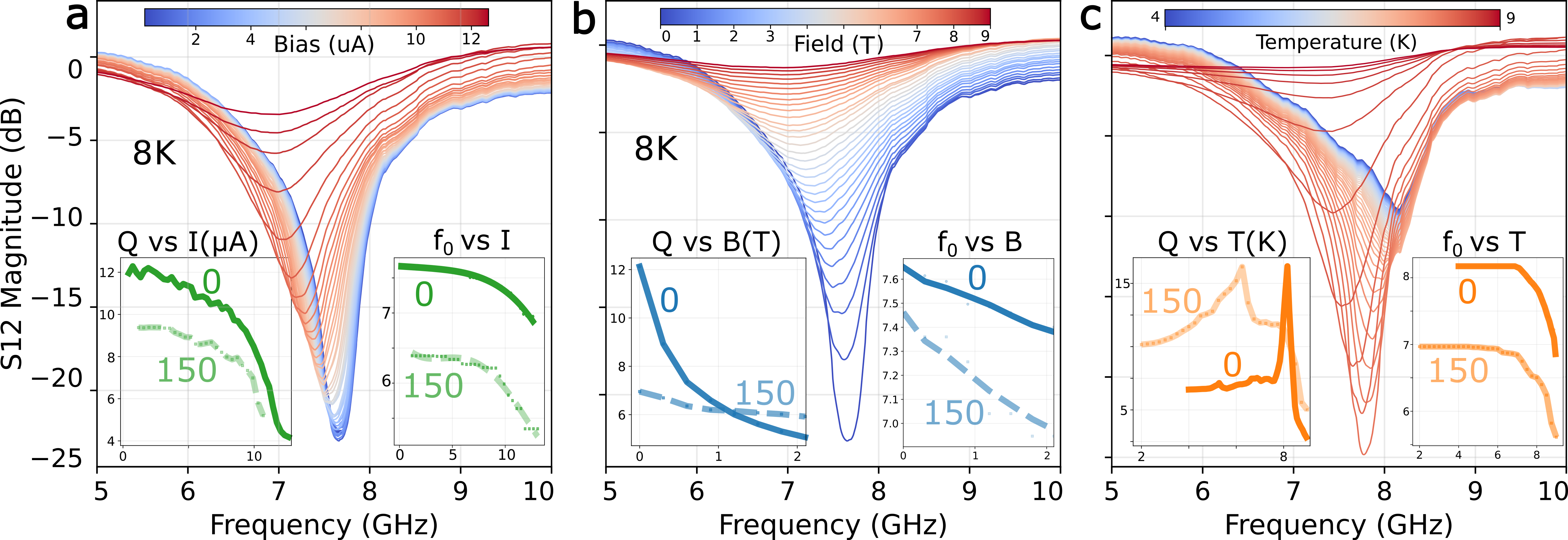}
    \caption{Depairing with bias current, magnetic field, and temperature. Microwave transmission $|S_{12}|$ spectra of the non-irradiated device as a function of (a) dc bias current at $8$~K, (b) perpendicular magnetic field at $8$~K, and (c) temperature. Increasing bias current, temperature or field shifts the resonance to lower frequency. The left and right insets show the extracted $Q$ and $f_0$ (in GHz) versus bias current (in $\mu\mathrm{A}$), field (in T), and temperature (in K), respectively, with axis limits adjusted to highlight optimal contrast between non-irradiated and 150~ions/nm$^{2}$ irradiated devices.
}
    \label{fig:fig5}
\end{figure*}

To explore superconducting depairing mechanisms, we compare the quality factor under three depairing knobs: temperature, field and bias current. As shown in Fig.~\ref{fig:fig5}, the measured $|S_{12}|$ spectrum scales very differently with each parameter because each parameter suppresses superconductivity through different microscopic channels. This enables a direct, mechanism-level comparison using $Q$ as a common observable, allowing the dominant microwave loss channel to be identified for each perturbation. 

Under DC bias current, $Q(I)$ exhibits a strongly nonlinear collapse (in Fig. 5a inset) that sharpens near the depairing current because the superfluid density is suppressed by current-induced pair breaking, which increases the kinetic inductance, $L_k \propto \left[1-(I/I_{\mathrm{dep}})^2\right]^{-1}$, implying a quadratic onset near $I_{\mathrm{dep}}$. As $I \rightarrow I_{\mathrm{dep}}$, the real part of the microwave surface impedance increases due to enhanced quasiparticle generation and nonlinear conductivity, leading to increased dissipation and a corresponding reduction of $Q$.

Magnetic-field-induced depairing results in a fundamentally different scaling of $Q$ in the Fig. 5b inset. In type-II superconducting thin films in perpendicular fields, the vortex areal density $n_v$ is given by the flux density divided by the flux quantum, $n_v = B/\Phi_0$. Microwave loss from vortices is commonly modeled by vortex dynamics~\cite{coffey1991unified}, where vortices add an effective complex resistivity $\propto n_v$. Notably, the SNSPDs exhibited no substantial change in behavior as a function of field at base temperature. No previous work has observed SNSPD operation at fields this large\cite{lawrie2021multifunctional}, and the ability to operate in 9T fields unlocks new opportunities for quantum sensing.

Temperature tuning reveals qualitatively distinct behavior, with $Q(T)$ (in Fig. 5c inset) exhibiting a pronounced maximum near $T_c$, indicative of a crossover between competing intrinsic dissipation mechanisms. At lower temperatures, microwave loss is weakly temperature dependent and can include contributions from two-level systems (TLSs), while near $T_c$, quasiparticle dissipation increases rapidly as thermal excitations grow~\cite{alexander2022power}. As $T$ approaches $T_c$, the sharp increase in thermally excited quasiparticles, described by Mattis--Bardeen electrodynamics, leads to a rapid collapse of $Q$~\cite{mattis1958theory}. 
In the irradiated device, the $Q(T)$ maximum is significantly broadened, consistent with disorder-induced broadening of the quasiparticle density of states, which smears the onset of quasiparticle dissipation over a wider temperature range rather than enhancing TLS loss. By contrast, no analogous peak is observed under bias current or magnetic field tuning, where current-induced pair breaking and vortex-related dissipation introduce additional loss channels that dominate the microwave response and obscure any TLS–quasiparticle crossover. Power-, frequency-, and temperature-dependent measurements as a function of helium-ion fluence are required to determine how disorder alters the dominant microwave dissipation mechanisms in future work.

By accessing intrinsic depairing physics, that is missed by conventional DC transport, dark-count-rate, and photoresponse measurements, our SNSPD microwave spectroscopies highlight a useful and complementary disorder probe. The large separation between the local switching current, dark-count onset, and depairing current reveals that SNSPD operation is governed by localized instabilities rather than the global superconducting limit. Local helium-ion irradiation enables controlled redistribution of disorder, selectively reducing the intrinsic depairing current while leaving pre-existing geometric bottlenecks largely unaffected, thereby closing the gap between $I_{sw}$ and $I_{dep}$. Beyond local manipulation of SNSPD operating parameters, our results establish microwave transmission as a spectroscopy of electrodynamic inhomogeneity, capable of resolving distinct superconducting regions within a single device. This framework provides a direct link between microscopic disorder, kinetic inductance nonlinearities, and macroscopic device performance, as well as a scalable strategy for disorder-engineered superconducting detectors and resonators.

\acknowledgments
This research was sponsored by the U. S. Department of Energy, Office of Science, Basic Energy Sciences, Materials Sciences and Engineering Division. Helium ion implantation was performed through a user project supported by the Center for Nanophase Materials Sciences (CNMS), which is a US Department of Energy, Office of Science User Facility at Oak Ridge National Laboratory.  E.D. was supported by the U.S. Department of Energy, Office of Science, Advanced Scientific Research Program, Early Career Award under Contract No. ERKJ420.

%

\end{document}